\documentclass[aps,prb,twocolumn,superscriptaddress]{revtex4-2}

\usepackage{amsmath}
\usepackage{mathtools} % successor of amsmath
\usepackage{amssymb}
\usepackage{amsthm}
\usepackage{bm,upgreek} % Allows use of bold-italic greek letters
\usepackage{upgreek} % roma greek
\usepackage{xcolor}
\usepackage{graphicx}
\usepackage{epstopdf}
\usepackage{ulem}
\usepackage[colorlinks,linkcolor=blue,anchorcolor=blue,citecolor=blue,urlcolor=blue]{hyperref}
\usepackage{ulem,cancel}
\newcommand{\nn}{{\nonumber}}
\newcommand{\bea}{\begin{eqnarray}}
\newcommand{\eea}{\end{eqnarray}}
\newcommand{\ie}{\textit{i.e.{ }}}

\newcommand{\up}{\uparrow}
\newcommand{\dn}{\downarrow}

\newcommand{\0}[1]{\mathbf{#1}}
\newcommand{\av}[1]{\left\langle #1 \right\rangle}
\newcommand{\bra}[1]{\left\langle #1 \right|}
\newcommand{\ket}[1]{\left| #1 \right\rangle}

\begin{document}

\title{Possible $S_\pm$-wave superconductivity in La$_3$Ni$_2$O$_7$}
\author{Qing-Geng Yang}
\affiliation{National Laboratory of Solid State Microstructures $\&$ School of Physics, Nanjing University, Nanjing, 210093, China}
%\author{Han-Yang Liu}
%\affiliation{National Laboratory of Solid State Microstructures $\&$ School of Physics, Nanjing University, Nanjing, 210093, China}
\author{Da Wang}\email{dawang@nju.edu.cn}
\affiliation{National Laboratory of Solid State Microstructures $\&$ School of Physics, Nanjing University, Nanjing, 210093, China}
\affiliation{Collaborative Innovation Center of Advanced Microstructures, Nanjing 210093, China}
\author{Qiang-Hua Wang} \email{qhwang@nju.edu.cn}
\affiliation{National Laboratory of Solid State Microstructures $\&$ School of Physics, Nanjing University, Nanjing, 210093, China}
\affiliation{Collaborative Innovation Center of Advanced Microstructures, Nanjing 210093, China}

\begin{abstract}
Recently, the bulk nickelate La$_3$Ni$_2$O$_7$ is reported to show signature of high-temperature superconductivity under high pressure above $14$GPa [\href{https://www.nature.com/articles/s41586-023-06408-7}{H. Sun {\it et al.}, Nature {\bf 621}, 493 (2023)}].
We analyze the pairing mechanism and pairing symmetry in a bilayer Hubbard model with two orbitals in the $E_g$ multiplet. In the weak to moderate interaction regime, our functional renormalization group (FRG) calculations yield $S_\pm$-wave Cooper pairing triggered by leading spin fluctuations. The gap function changes sign across the Fermi pockets, and in real space the pairing is dominated by intra-unitcell intra-orbital components with antiphase between the onsite ones. In the strong coupling limit, we develop a low-energy effective theory in terms of atomic one- and two-electron states in the $E_g$ multiplet. The variational treatment of the effective theory produces results consistent with the FRG ones, suggesting the robustness of such a pairing function. The driving force for superconductivity in the strong coupling limit can be attributed to the local pair-hopping term and the spin-exchange on vertical bonds. We also discuss a possible scenario for the weak insulating behavior under low pressures in terms of the tendency toward the formation of charge order in the strong coupling limit.
\end{abstract}
\maketitle

{\it Introduction}.
After the discovery of cuprate high-$T_c$ superconductors \cite{cuprate_1986,Lee_RMP_2006}, there are continuing efforts to search for superconductivity in perovskites of similar structure, such as ruthenates \cite{Maeno_N_1994},
cobaltates \cite{cobaltate_2003} and nickelates \cite{Li_N_2019}, in order to understand how unique the copper is in cuprates on one hand, and to discover new families of high-$T_c$ superconductors on the other hand. As a possible breakthrough along this line, a recent report shows signature of superconductivity in the bulk nickelate La$_3$Ni$_2$O$_7$ with $T_c$ as high as $80K$ at a pressure above $14$GPa \cite{exp}.
It has a bilayer perovskite structure with the nickel atom sitting in the center of each oxygen octahedron, with layered NiO$_2$-plane similar to the CuO$_2$-plane in cuprates. However, the valence of Ni-atom fluctuates between Ni$^{2+}$ and Ni$^{3+}$, with an average $3d^{7.5}$ atomic configuration. Under high pressure,
both $3d_{x^2-y^2}$- and $3d_{3z^2-r^2}$-orbitals in the $E_g$-multiplets are active near the Fermi level \cite{exp,model}.
The $E_g$ multiplet in La$_3$Ni$_2$O$_7$ is also different to the $t_{2g}$ multiplet in iron-based superconductors \cite{Fe-based_2008,Stewart_RMP_2011} where the Fe-atom is in an edge-sharing tetrahedron.
At the fractional filling, La$_3$Ni$_2$O$_7$ is unlikely to form
a Mott insulator. But quite unexpectedly, the material is weakly insulating at lower pressures \cite{exp}.

The novel attributes of La$_3$Ni$_2$O$_7$ motivate us to explore the pairing mechanism and pairing symmetry therein, and to understand the weak insulating behavior under lower pressures. We perform the analysis
within a bilayer two-orbital Hubbard model by functional renormalization group (FRG) in the the weak to moderate interaction regime, and effective theory in the strong coupling limit. We find the Cooper pairing has $S_\pm$-wave symmetry. In momentum space, the sign change of the gap function is compatible with the structure of leading spin fluctuations, suggesting pairing triggered by spin fluctuations. In real space, the Cooper pair is dominated by intra-orbital pairing on-site and on the vertical inter-layer bond, with stronger amplitudes in the $d_{3z^2-r^2}$-orbital content, anti-phase between the on-site pairs. In the strong coupling limit, we develop a low-energy effective theory in terms of atomic one- and two-electron states in the $E_g$ multiplet. The results from variational treatment of the effective theory are consistent with the FRG results. The driving force for superconductivity in the strong coupling limit can be attributed to the local pair-hopping interaction and the spin-exchange of $d_{3z^2-r^2}$-electrons on vertical bonds. We also discuss a possible scenario of the weak insulating behavior under lower pressures in terms of the tendency toward charge ordering in the strong coupling limit.

{\it Model}.
We start from a two-orbital ($d_{3z^2-r^2}$ and $d_{x^2-y^2}$) tight-binding model on a bilayer square lattice as extracted from the Wannier fitting to the band structure of density functional theory (DFT) \cite{model},
\begin{align}
H_0=\sum_{i\delta,ab,\sigma}t_{\delta}^{ab}c_{ia\sigma}^\dag c_{i+\delta b\sigma} + \sum_{ia\sigma} \varepsilon_a c_{ia\sigma}^\dag c_{ia\sigma} ,
\end{align}
where $t_{\delta}^{ab}$ is the hopping matrix element between $a$-orbital on site $i$ and $b$-orbital on site $i+\delta$, $\sigma$ denotes spin, $\varepsilon_a$ is the on-site energy of the $a$-orbital. Up to $C_{4v}$ symmetry, the tight-binding parameters are given by  ($t^{xx}_{100}$, $t^{xx}_{110}$, $t^{xx}_{00\frac12}$, $t^{zz}_{100}$, $t^{zz}_{110}$, $t^{zz}_{00\frac12}$, $t^{xz}_{100}$, $t^{xz}_{10\frac12}$, $\varepsilon_{x}$, $\varepsilon_{z}$)$=$($-0.483$, $0.069$, $0.005$, $-0.110$, $-0.017$, $-0.635$, $0.239$, $-0.034$, $0.776$, $0.409$)~eV \cite{model}. Here $x$/$z$ denotes the $d_{x^2-y^2}$/$d_{3z^2-r^2}$ orbital, and the vertical inter-layer distance is assigned as $\frac12$. We note that the hopping of $d_{3z^2-r^2}$-electron along the vertical inter-layer bond is the strongest.

\begin{figure}
\includegraphics[width=\linewidth]{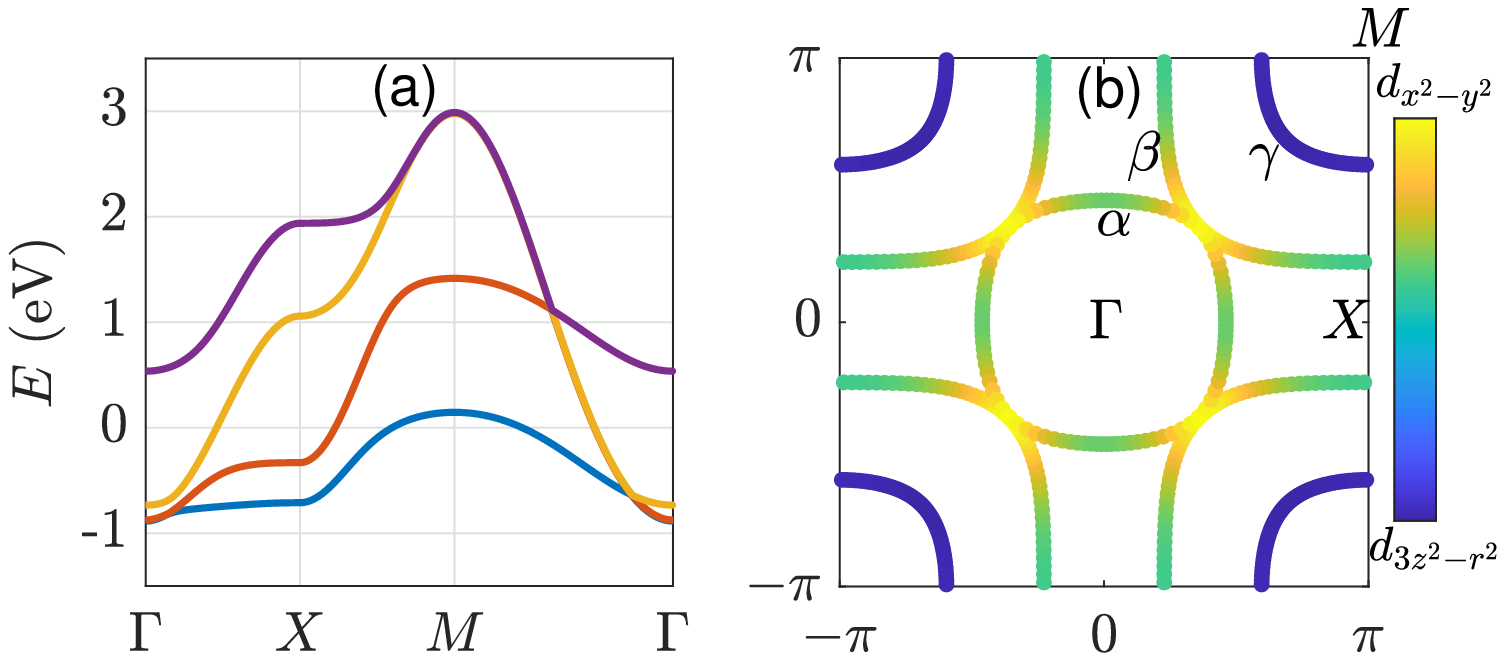}
\caption{\label{fig:band}
(a) Band dispersion along the high symmetry lines $\Gamma-X-M-\Gamma$. (b) Fermi surfaces in the Brillouin zone with color scaled orbital characters. The three pockets are labeled by $\alpha$, $\beta$ and $\gamma$, respectively.
}
\end{figure}

The energy bands (a) and Fermi surfaces (b) are shown in Fig.~\ref{fig:band}. There are three bands crossing the Fermi level, leading to three Fermi pockets: one electron-like $\alpha$-pocket around $\Gamma$, one small hole-like $\gamma$-pocket around $M$, and another large hole-like $\beta$-pocket around $M$. From the color scale in Fig.~\ref{fig:band}(b), both the $\alpha$- and $\beta$-pockets show strong orbital hybridizations, while the $\gamma$-pocket mainly comes from the $d_{3z^2-r^2}$-orbital. From the DFT calculations, the shallow $\gamma$-pocket only appears in the high pressure phase, while it vanishes in the low pressure phase \cite{exp}, suggesting its importance for superconductivity in this material.

We consider the multi-orbital atomic Coulomb interactions
\begin{align}
H_I&=\frac12\sum_{i,a\ne b,\sigma\sigma'}\left( U'n_{ia\sigma}n_{ib\sigma'}+J_Hc_{ia\sigma}^\dag c_{ib\sigma} c_{ib\sigma'}^\dag c_{ia\sigma'} \right)\nonumber\\
&+\sum_{ia} Un_{ia\up}n_{ia\dn}+\sum_{i,a\ne b} J_Pc_{ia\up}^\dag c_{ia\dn}^\dag c_{ib\dn} c_{ib\up},
\end{align}
where $U$ is the intra-orbital Hubbard repulsion, $U'$ is the inter-orbital Coulomb interaction, $J_H$ is the Hund's coupling, and $J_P$ is the pair hopping interaction. Throughout this work, we use the Kanamori relations \cite{KanamoriRelations} $U=U'+2J_H$ and $J_H=J_P$.

{\it Functional renormalization group}.
We first perform the singular-mode functional renormalization group (SM-FRG) \cite{Wang_PRB_2012, Xiang_PRB_2012} calculations to look for possible electronic instabilities in the system. The FRG approach is controlled in the weak to moderate interaction regime \cite{Honerkamp_PRB_2001,Honerkamp_PRB_2004,Metzner_RMP_2012,Yamase_PRL_2016}.
It provides the flow of the effective interactions between quasi-particles versus a running infrared cutoff energy scale $\Lambda$ (the lowest Matsubara frequency in our case). In our SM-FRG, we define a fermion bilinear $c_{a\sigma}^\dagger (\0R)c_{b\sigma'}^\dagger(\0R+\delta)$ in the pairing channel, labelled by the orbital-spin combination $(a\sigma,b\sigma')$ and the internal displacement $\delta$, in addition to the shared position $\0R$. Similar bilinears can be defined in the particle-hole channels, and are transformed into the momentum space in the calculations. As $\Lambda$ is lowered from the ultra-violet limit, we extract from the full effective interactions the scattering matrices $V_{\rm SC,SDW,CDW}(\0q)$ between fermion bilinears (of momentum $\0q$) in the superconducting (SC), spin-density-wave (SDW) and charge-density-wave (CDW) channels, respectively.
We monitor the flow of the leading negative eigenvalue $S$ in each channel. The first divergence at a critical scale $\Lambda=\Lambda_c$ signals an emerging electronic order at transition temperature $T_c\sim \Lambda_c$, described by the bilinear combination in the eigen scattering mode of the diverging channel. The technical details of our SM-FRG can be found in {Supplementary Materials} \cite{sm} and Refs.~\cite{Wang_PRB_2012,Wang_PRB_2013,Tang_PRB_2019}.

\begin{figure}
\includegraphics[width=\linewidth]{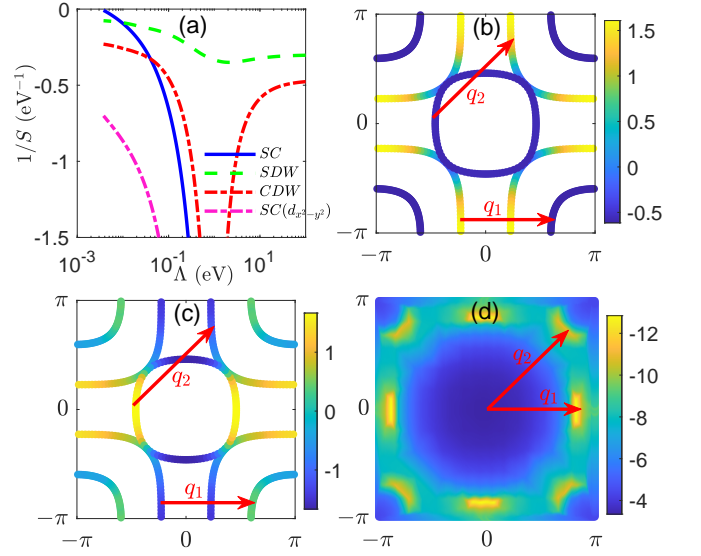}
\caption{\label{fig:flow-sc}
(a) FRG flow of the leading eigenvalues $S$ (plot as $1/S$) versus the running energy scale $\Lambda$ in the SC, SDW and CDW channels, for $(U,J_H)=(3,0.3)$~eV. For comparison, the subleading eigenvalue in the pairing channel is also plotted. (b) and (c) show the leading $s$-wave and subleading $d_{x^2-y^2}$-wave gap functions on the Fermi surfaces (color scale). (d) shows the renormalized interaction $V_{\rm SDW}(\0q)$ when the SC channel diverges. The two arrows indicate two dominant scattering momenta $\0q_{1,2}$, which connect Fermi points with opposite gap signs as seen in (b) and (c).
}
\end{figure}

A typical FRG flow is shown in Fig.~\ref{fig:flow-sc}(a), where we plot $1/S$ versus $\Lambda$ for the case of $(U,J_H)=(3,0.3)$~eV. The SDW channel is strongest at high energy scales, as in the bare interaction. The CDW channel is initially reduced by Coulomb screening, which also causes the SDW channel to decrease slightly. The SC channel does not show up initially since it is repulsive. The SDW channel begins to increase further as $\Lambda\sim 1$~eV, as off-site spin correlations are induced. Meanwhile, through the channel overlap, the SDW channel triggers the rise of the SC channel. This is a manifestation of the well-known Kohn-Luttinger mechanism \cite{Kohn-Luttinger_1965}. The CDW also rises because off-site bilinears (such as the valence bond) in this channel can be induced by SDW. As $\Lambda$ is lowered further, the SDW and CDW channels tend to saturate because the phase space for low-energy particle-hole excitations decreases. However, the SC channel at zero bilinear momentum $\0q = 0$ can continue to rise through the Cooper mechanism. Finally the SC channel diverges first, suggesting an instability of the normal state toward the SC state.

\begin{figure}
\includegraphics[width=\linewidth]{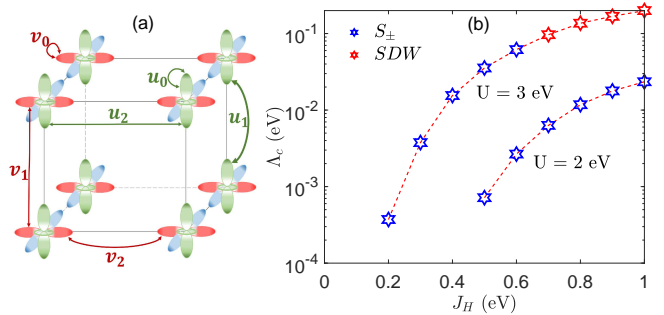}
\caption{\label{fig:phase} (a) Schematic plot of the dominant intra-orbital pairing components in real space, with $u_{0,1,2}$ for $d_{3z^2-r^2}$-orbital (vertical rods) and $v_{0,1,2}$ for $d_{x^2-y^2}$-orbitals (planar crosses), up to translation and $C_{4v}$ symmetry. (b) The critical scale $\Lambda_c$ versus $J_H$ for two values of $U$. Blue and red stars stand for $S_\pm$-wave SC and SDW, respectively.}
\end{figure}

We next analyze the pairing function. It can be obtained directly from the leading eigen mode of the SC-channel interaction $V_{\rm SC}(\0q=\00)$ in the fermion bilinear basis, see the Supplementary Material \cite{sm} for more details. The eigen mode corresponds to a pairing term $H_{SC}=\sum_{i\delta,ab}\Delta_{\delta}^{ab} c_{ia\up}^\dag c_{i+\delta b\dn}^\dag$, and the dominant pairing components are illustrated in Fig.~\ref{fig:phase}(a), where $u_{0,1,2}$ denote $\Delta_{\delta}^{zz}$,  and $v_{0,1,2}$ denote $\Delta_{\delta}^{xx}$.
In the case of $(U,J_H)=(3,0.3)$~eV and up to a global scale, (i) the pairing within the $d_{3z^2-r^2}$-orbital is $u_0=-0.43$ (onsite) and $u_1=0.51$ (on vertical bond). (ii) the pairing between the $d_{x^2-y^2}$-orbitals is $v_0=0.14$ (onsite) and $v_1=0.07$ (on vertical bond), (iii) the intra-orbital pairing on in-plane nearest-neighbor bonds are $(u_2, v_2)\sim (-0.05, -0.02)$ for $d_{3z^2-r^2}$ and $d_{x^2-y^2}$ orbitals, respectively.
The pairing between the $d_{3z^2-r^2}$-orbitals are much stronger. The anti-phase between $u_0$ and $v_0$ is consistent with the repulsive bare local pair-hopping interaction $J_P$. The pairing symmetry is clearly $s$-wave.

To gain more insights, we further project the pairing function into the band basis to obtain the gap function $\Delta_{n\0k}=\langle n\0k| \sum_{\delta}\Delta_\delta e^{i\0k\cdot\bm{\delta}} |n\bar{\0k}\rangle$ where $|n\0k\rangle$ is the Bloch state for band $n$, $\Delta_\delta$ is understood as a matrix in the orbital-sublattice basis, and we have used the time-reversal symmetry $|n\bar{\0k}\rangle = T|n\0k\rangle$ (for $\bar{\0k} = -\0k$) to simplify the expression. The gap function (color scale) is shown on the Fermi surfaces in Fig.~\ref{fig:flow-sc}(b).
Both $\alpha$- and $\gamma$-pockets are fully gapped with the same sign, while the gap on the $\beta$-pocket is largely of opposite sign and  diminishes in the nodal direction where the two pockets are very close. Therefore, exactly speaking, the pairing symmetry is $S_\pm$-wave, similar to the case in iron pnictides \cite{Chubukov_ARCMP_2012}, bilayer Hubbard models \cite{LeeDH,Scalapino2011,Kuroki2017}, and infinite layer Nickelates \cite{Eremin2022}.
On the other hand, we note that as the leading eigen mode diverges, the subleading one with $d_{x^2-y^2}$-pairing is finite, see Fig.\ref{fig:flow-sc}(a), and can be ruled out as a competitor for the SC state. However, for comparison, we also plot its gap function in Fig.~\ref{fig:flow-sc}(c). Since the $\gamma$-pocket is near $M$, the $d_{x^2-y^2}$-wave gap function is nodal, and tiny on the entire $\gamma$ pocket. This should be unfavorable to gain condensation energy if such a pairing would occur in the SC state. Interestingly, the experiment shows that the SC phase appears when the $\gamma$-pocket appears \cite{exp}. This fact supports our $S_\pm$-wave instead of $d_{x^2-y^2}$-wave as the leading pairing symmetry.

In Fig.~\ref{fig:flow-sc}(d), we plot the leading negative eigenvalue of the renormalized $V_{\rm SDW}(\0q)$ in the momentum space as the pairing channel diverges. We see the spin interaction is strong near momenta $\0q_1\sim (0.75, 0.75)\pi$ and $\0q_2\sim (0.84,0)\pi$, up to symmetry related images, which appear to connect the Fermi points where the gap function has opposite signs, as seen from the same arrows in Figs.~\ref{fig:flow-sc}(b) and \ref{fig:flow-sc}(c). This is consistent with pairing triggered by spin fluctuations, as the FRG flow already suggests. Similar mechanism is believed to work in cuprates and iron pnictides, at least in the weak-coupling scenario \cite{Scalapino_RMP_2012}.

To see the robustness of the $S_\pm$-wave pairing, we performed systematic calculations by varying $J_H$ and $U$, with $J_H\leq U/2$. Fig.~\ref{fig:phase}(b) shows the phase diagram. We find the $S_\pm$-wave pairing is always favored, and the critical scale increases with $J_H$, until the SDW channel begins to win for larger $U$ and $J_H$. The reason is the SDW interaction is enhanced by $J_H$, even at the bare level. While the exact values for $U$ and $J_H$ are unavailable at this stage, our results show that the SC phase is realized in a large regime of the parameter space.

\begin{figure}
    \includegraphics[width=\linewidth]{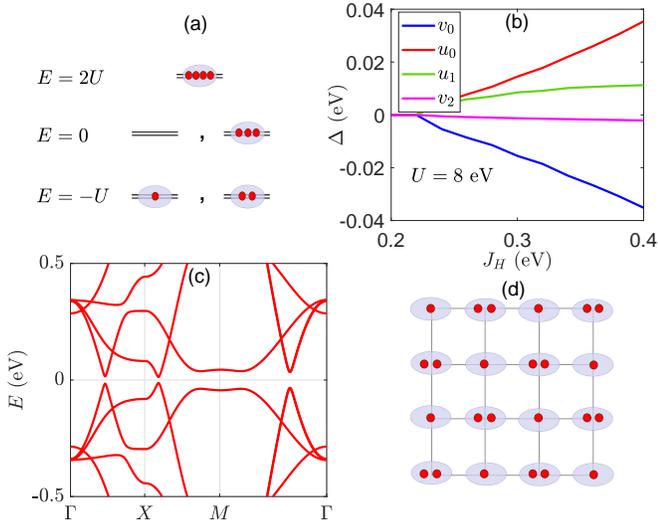}
    \caption{\label{fig:strong-coupling}
    (a) Atomic charge states and energy levels for $3/2$-filling. (b) Paring components in $H_v$ versus $J_H$ at $U=8$~eV. (c) Band structure using $H_v$ for $(U,J_H)=(8,0.3)$~eV. (d) A schematic plot of the possible in-plane CDW at lower pressures.}
    \end{figure}

{\it Strong coupling limit}.
We first assume $U=U'$, $J_H=J_P=0$, and ignore the small crystal field splitting and the kinetic hopping. (The difference to this situation will be taken as perturbations.) In the atomic limit, the one-electron (1e) and two-electron (2e) states on a site are degenerate if the chemical potential is set at $\mu=U$. The energy levels for all possible atomic charge states is illustrated in Fig.~\ref{fig:strong-coupling}(a). The ground state manifold is spanned by the 1e- and 2e-states, and the density of both atomic states should be 1/2 to match the average electron filling level $3/2$. Then up to the second order in the small perturbations, the effective Hamiltonian in the low energy sector can be written as,
\begin{align} \tilde{H}&=\sum_{\langle ij\rangle a,b,\sigma} \left(t_{ij}^{ab}c_{ia\sigma}^\dagger c_{jb\sigma}Q_{1e,i} Q_{2e,j}+{\rm h.c.}\right)\nonumber\\
&+\sum_{\langle ij\rangle a,n=1,2}J_{ij}^a \left({\bf S}_{ia}\cdot {\bf S}_{ja}-\frac{1}{4}n_{ia}n_{ja}\right)Q_{ne,i}Q_{ne,j}\nonumber\\
&+\sum_{i,a}J_H\left(Q_{2e,ia}-Q_{1e,ia}Q_{1e,i\bar{a}}\right)\nonumber\\
&+\sum_{i,a>b}J_H\left(\frac{1}{4}n_{ia}n_{ib}-{\bf S}_{ia}\cdot {\bf S}_{ib}\right)Q_{1e,ia}Q_{1e,ib}\nonumber\\
&+\sum_{i,a>b}J_P\left(c_{ia\uparrow}^\dagger c_{ia\downarrow}^\dagger c_{ib\downarrow}c_{ib\uparrow}+{\rm h.c.} \right)\nonumber\\
&+\sum_{ia} \epsilon_a\left(Q_{1e,ia}+2Q_{2e,ia}\right). \label{eq:heff}
\end{align}
This can be taken as a generalized two-orbital $t-J$ model, but with the important difference that double occupancy is allowed. To be more specific,
for a given site $i$, $Q_{ne,ia}$ is a projection operator for charge $ne$ on the $a$-orbital, ${\bf S}_{ia}$ for spin operator, and $Q_{ne,i}$ for total charge $ne$ at site $i$. The second line is the super-exchange by second-order hopping processes via an intermediate atomic 3e- or 0e-state of energy $\bar{U}\sim (U+U')/2=U-J_H$, resulting in the antiferromagnetic coupling $J_{ij}^a\sim 4(t_{ij}^{aa})^2/\bar{U}$.
Note we ignored super-exchange processes involving $t_{ij}^{ab}$ or $t_{ij}^{aa} t_{ji}^{bb}$ with $a\neq b$, since they are much smaller than the leading orbital-diagonal cases. In the third line, $\bar{a}$ denotes the orbital other than $a$, and in the last line we subtracted the average of the crystal field to write $\epsilon_a = \varepsilon_a - \bar{\varepsilon}$. On the other hand, we dropped a constant $U$ (per site), which is the largest energy scale we designed to eliminate. The Hamiltomian acts on a many-body state under a global projection operator $P = P_{N_e} \Pi_i (Q_{1e,i}+Q_{2e,i})$, where $P_{N_e}$ is a projection operator for the total charge, $\sum_{i}(Q_{1e,i} + 2 Q_{2e,i}) = N_e$.

As the usual one-band $t$-$J$ model, the above model still contains strong correlation effects due to the projection operators and constrains. We proceed to investigate the (uniform paramagnetic) ground state $|G\rangle $ by variational principle: $|G\rangle = P |0\rangle$, where $|0\rangle$
is the ground state of a variational free Hamiltonian,
\begin{align} H_v &= \sum_{ ijab\sigma} \left[h_{ ij}^{ab}c_{ia\sigma}^\dagger c_{ja\sigma}+\sigma \left(\Delta_{ij}^{ab}c_{ia\sigma}^\dagger c_{jb\bar{\sigma}}^\dagger+{\rm h.c.}\right)\right],\nonumber
\end{align}
where $h$'s describe hopping on bonds as well as onsite energy and orbital hybridization, and $\Delta$'s are the corresponding singlet pairing amplitudes. The energy $E = \langle G|\tilde{H}|G\rangle /\langle G|G\rangle$ needs to be minimized by optimizing the variational parameters in $H_v$. This could be performed by variational Monte Carlo, but for a flavor of the underlying physics, we perform the minimization semi-analytically in the Gutzwiller approximation \cite{ZhangFC1988,Wang_PRB_2006}. See Supplemental Material \cite{sm} for details.
The variational calculation always ends up with intra-orbital $s$-wave singlet pairing, with dominant real-space components as schematically shown in Fig.~\ref{fig:phase} (a). The numerical value is shown in Fig.~\ref{fig:strong-coupling}(b) for these components versus $J_H$ for $U=8$~eV. The two onsite intra-orbital components $u_0$ and $v_0$ are large and always anti-phase, increasing with $J_H$, the component $u_1$ on vertical bond for $d_{3z^2-r^2}$-orbital is sub-leading, while the other components are much weaker. This demonstrates that our FRG results obtained for weaker interactions are robust and suggestive even for the strong coupling case. Moreover, we find if we switch off the $J_P$-term in Eq.~\ref{eq:heff}, then $(u_0, v_0)\to 0$ and only $u_1$ survives and is also small. This shows that the driving force for pairing in the strong coupling limit is the local pair-hopping interaction and spin-exchange of $d_{3z^2-r^2}$-electrons on vertical bonds. Fig.~\ref{fig:strong-coupling}(c) shows the Bogoliubov-de Gennes band structure using $H_v$. Clearly, the gap opens along the entire line cuts, consistent with the $s$-wave symmetry. We also checked that the normal state part of $H_v$ leads to a bandwidth that is roughly a half of the unprojected one. We should point out the above theory applies best for smaller values of $J_H$. A larger $J_H$ may eventually drive the system into magnetic ground states \cite{Dagotto-arxiv}.

Finally, we recall that in a usual one-band Hubbard model on the square lattice, the nearest-neighbor Coulomb interaction $V$ would favor a charge ordered state provided that $V>U/4$ \cite{Yao_PRB_2022}. However, in the above projected Hamiltonian suitable for the average $3/2$-filling, the 1e- and 2e-states are near degenerate at the atomic level, so that $U$ does not hamper the charge ordering of the 1e- and 2e-states. In fact, it is advantageous to form a staggered charge order of 1e- and 2e-states on the lattice, a form of Wigner crystal, see Fig.~\ref{fig:strong-coupling}(e), to save Coulomb energy. The charge order only has to fight against the kinetic quantum fluctuations, and would set in once $4V\chi_c\geq 1$ in the simple Stoner picture, where $\chi_c$ is the charge susceptibility. In the present case, the projection reduces the bandwidth and enhances the charge susceptibility. More interestingly, at lower pressures the bare bandwidth is smaller, so that the charge order is more likely to occur. The charge order reduces the charge mobility and the transition may be first order. This may explain the weak insulating behavior at lower pressures, and the abrupt emergence of the SC state at higher pressures \cite{exp}, in addition to the effect of possible aptical oxygen vacancies.

%%%%%%%%%%%%%%%%%%%%%%%%%%%%%%%%%%%%%%%%%%
%{\it Note added}. After submission, we become aware of two related works %\cite{Dagotto-arxiv,Eremin-arxiv}.

{\it Acknowledgement}.
This work is supported by National Key R\&D Program of China (Grant No. 2022YFA1403201) and National Natural Science Foundation of China (Grant No. 12374147, 12274205, and 11874205).

\bibliography{La3Ni2O7}

%%%%%%%%%%%%%%%%%%% supplementary
\appendix
\begin{widetext}

\vspace{10pt}
\begin{center}{\Large \textbf{
Supplementary Material
}}\end{center}
%{Supplementary Material on ``Possible $S_\pm$-wave superconductivity in La$_3$Ni$_2$O$_7$''}
%\author{Qing-Geng Yang}
%\affiliation{National Laboratory of Solid State Microstructures $\&$ School of Physics, Nanjing University, Nanjing, 210093, China}
%\author{Da Wang}\email{dawang@nju.edu.cn}
%\affiliation{National Laboratory of Solid State Microstructures $\&$ School of Physics, Nanjing University, Nanjing, 210093, China}
%\affiliation{Collaborative Innovation Center of Advanced Microstructures, Nanjing 210093, China}
%\author{Qiang-Hua Wang} \email{qhwang@nju.edu.cn}
%\affiliation{National Laboratory of Solid State Microstructures $\&$ School of Physics, Nanjing University, Nanjing, 210093, China}
%\affiliation{Collaborative Innovation Center of Advanced Microstructures, Nanjing 210093, China}
%\begin{abstract}

This Supplementary Material is divided into two parts. In the first part, we give an introduction of the singular-mode functional renormalization group (SM-FRG) method and provide a benchmark of the fermion bilinear truncation approximation, which is found to be very efficient in this problem. In the second part, we apply the Gutzwiller approximation to the effective Hamiltonian obtained in the strong-coupling limit, which gives the same pairing symmetry as obtained by SM-FRG in the weak-coupling regime.
%\end{abstract}
%\maketitle

\section{Singular-mode functional renormalization group}
As one of the powerful methods to study correlated electronic systems, the functional renormalization group (FRG) has been well explained in many documents \cite{Berges_PR_2002,Metzner_RMP_2012,Dupuis_PR_2021,Kopietz__2010}. In this appendix, we mainly focus on one of its realizations, called singular-mode FRG (SM-FRG).

\subsection{Mandelstam parameterization and 1PI flow equations}
In our SM-FRG, we study the FRG flows of the 4-point one-particle irreducible (1PI) vertices $\Gamma_{1234}$ appearing in the effective interaction  $H_{\Gamma}=\frac{1}{2}\sum_{1,2,3,4}\psi^\dagger_1\psi^\dagger_2\Gamma_{1234}\psi_3\psi_4$, where $\psi$ is the fermion field with the subscript $1,2,3,4$ denoting the one-particle degrees of freedom (such as frequency, momentum, orbital, sublattice, spin, etc.). In the SU(2) symmetric case under concern, the spins for 1 and 4 are the same, and similarly for 2 and 3.

\begin{figure} [h]
	\includegraphics[width=0.4\textwidth]{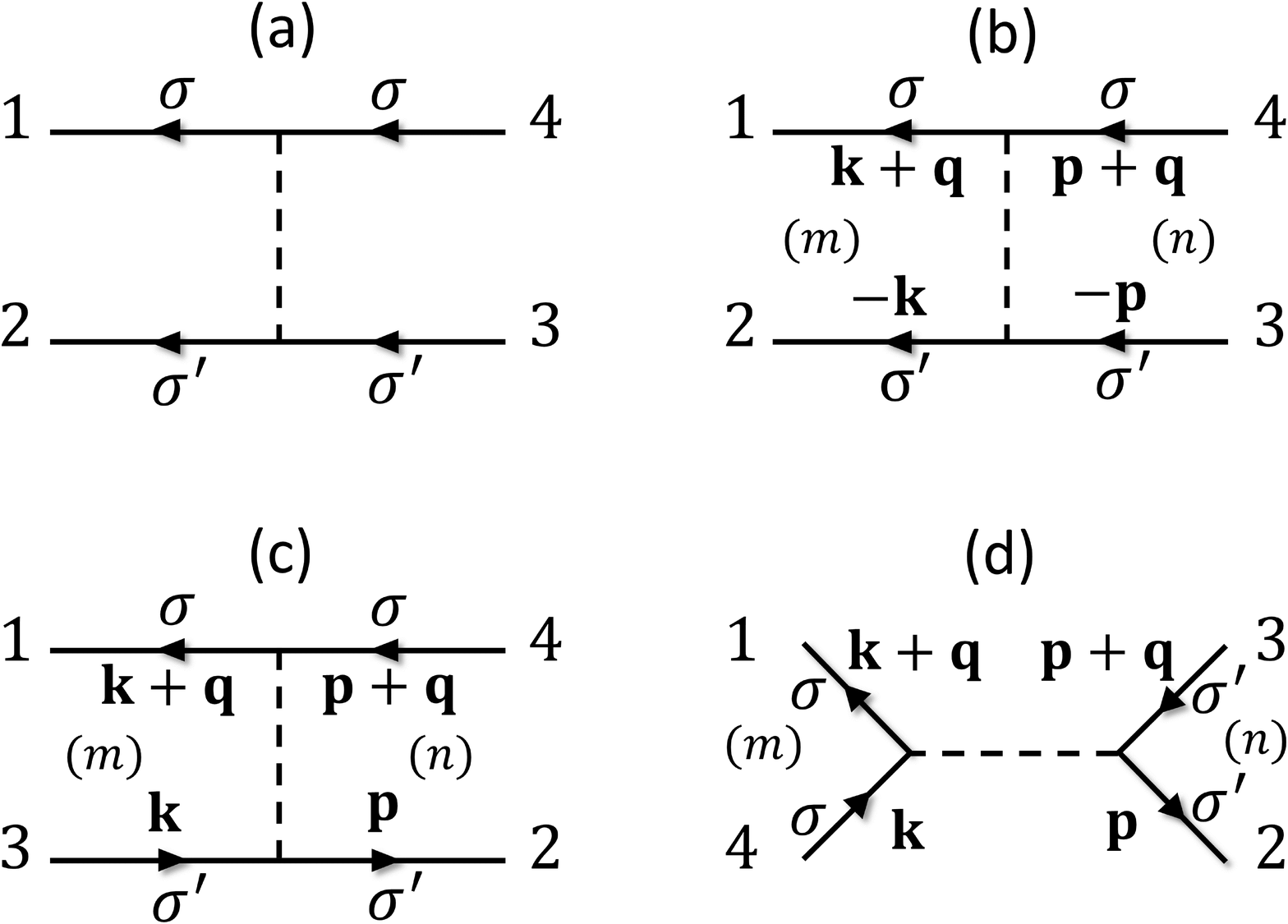}
	\caption{A generic 4-point 1PI vertex (a) can be rearranged into the pairing (P), crossing (C) and direct (D) channels as shown in (b)-(d), respectively.
	The momentum $\mathbf{k,q,p}$ are explicitly shown for clarity. The spins ($\sigma$ and $\sigma'$) are conserved during fermion propagation in the spin-SU(2) symmetric case. The labels $m$ and $n$ denote fermion bilinears.}
	\label{fig:vertex}
\end{figure}

We define fermion bilinears in the three Mandelstam channels as,
\begin{align}
\alpha_{12}^{\dag}=\psi_1^\dag\psi_2^\dag \ \text{ (pairing)}, \quad \beta_{13}^{\dag}=\psi_1^\dag\psi_3\  \text{ (crossing)}, \quad \gamma_{14}^{\dag}=\psi_1^\dag\psi_4 \ \text{ (direct)}.
\end{align}
Then the general 1PI vertex can be rewritten as scattering matrices $P$, $C$ and $D$ in the three channels as
\begin{align}
H_\Gamma=\frac12 \sum_{12,43}\alpha_{12}^{\dag} ~P_{12;43}~\alpha_{43} = -\frac12 \sum_{13,42}\beta_{13}^{\dag} ~C_{13;42}~ \beta_{42} = \frac12 \sum_{14,32}\gamma_{14}^{\dag} ~D_{14;32} ~\gamma_{32}, \label{eq:rewind}
\end{align}
as illustrated in Fig.~\ref{fig:vertex}(b)-(d).
The collective 4-momentum (between the two fermion bilinears) is
$q = k_1+k_2$, $k_1-k_3$, $k_1-k_4$, in the P, C and D channels.
If the subscripts $1,2,3,4$ run over all sites, the three scattering matrices $P$, $C$ and $D$ are all equivalent to $\Gamma$, i.e. $\Gamma_{1234}=P_{12;43} =C_{13;42}=D_{14;32}$.
But in practical calculations, the fermion bilinears must be truncated. On physical grounds, the important bilinears are those that join the singular scattering modes, and such eigen modes determine the emerging order parameter. Since order parameters are composed of short-ranged bilinears, only such bilinears are important. These include onsite and on-bond pairing in the pairing channel, and onsite and on-bond particle-hole density in the C and D channels. The FRG based on the decomposition of the interaction vertices into scattering matrices in the truncated fermion bilinear basis, which are sufficient to capture the most singular scattering modes, is called the singular-mode FRG (SM-FRG). \cite{Wang_PRB_2012,Xiang_PRB_2012,Wang_PRB_2013}

\begin{figure} [h]
	\includegraphics[width=0.4\textwidth]{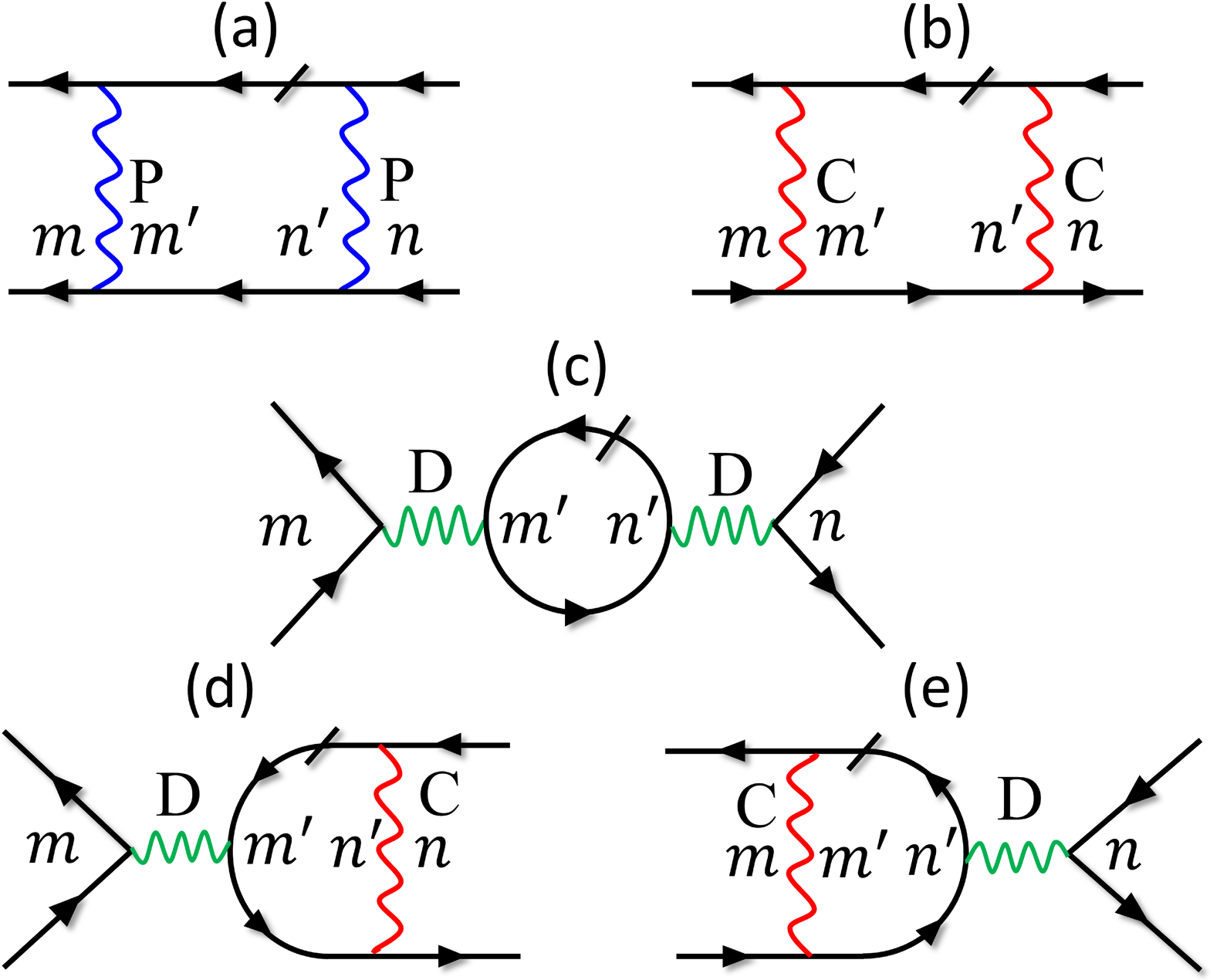}
	\caption{ One-loop contributions to $\partial\Gamma_{1234}/\partial\Lambda$. The wavy lines denote the truncated $P$ (blue), $C$ (red) and $D$ (green) from $\Gamma$. The slash in each diagram denotes the single-scale propagator and is put on either one of the fermion lines within the loop. The symbols $m$ and $n$ denote the fermion bilinear in the respective channels, and the inner indices are summed implicitly.}
	\label{fig:1loop}
\end{figure}

Starting from $\Lambda=\infty$ where the 1PI vertices $P$, $C$ and $D$ are given by the bare interactions, $\Gamma_{1234}$ flows as
\begin{align} \label{eq:flow}
\frac{\partial\Gamma_{1234}}{\partial\Lambda} &= [P\chi_{pp}P]_{12;43}+[C\chi_{ph}C]_{13;42} + [D\chi_{ph}C+C\chi_{ph}D-2D\chi_{ph}D]_{14;32} ,
\end{align}
see Fig.~\ref{fig:1loop} for illustration.
The products within the square brackets imply matrix convolutions, and $\chi_{pp}$ and $\chi_{ph}$ are single-scale (at $\Lambda$) susceptibilities given by, in real space,
\begin{align}
	[\chi_{pp}]_{ab;cd} &= \frac{1}{2\pi}\left[ G_{ac}(i\Lambda)G_{bd}(-i\Lambda)+
	(\Lambda \to -\Lambda) \right] \\
	[\chi_{ph}]_{ab;cd} &= \frac{1}{2\pi}\left[ G_{ac}(i\Lambda)G_{db}(i\Lambda)+
	(\Lambda \to -\Lambda) \right],
\end{align}
where $a,b,c,d$ are dummy fermion indices (that enter the fermion bilinear labels), and $G_{ab}(i\Lambda)$ is the normal state Matsubara Green's function. (The expression in the momentum space is slightly more complicated but is otherwise straightforward, and is used in actual calculations.)
As usual \cite{Metzner_RMP_2012}, we neglect self-energy correction which could be absorbed in the band dispersion, and we also neglect the sixth and higher order vertices, which are RG irrelevant. The frequency dependence of the 4-point vertices is also RG irrelevant and ignored. In this spirit, all external legs are set at zero frequency.
The functional flow equation is solved by numerical integration over $\Lambda$.
Note that after $\Gamma_{1234}$ is updated after an integration step, it is rewinded as $P$, $C$ and $D$ according to $\Gamma_{1234}=P_{12;43} =C_{13;42}=D_{14;32}$, subject to truncation of the fermion bilinears to be discussed below.
In this way, SM-FRG can treat interactions in all channels on equal footing. In fact, if we ignore the channel overlaps, the flow equation reduces to the ladder equation in the P-channel, and the random-phase-approximation in the C- and D-channels. The SM-FRG combines the three channels coherently.

\subsection{Singular scattering modes and order parameters}
The scattering matrices in the SC, SDW and CDW channels can be shown to be related to $P$, $C$ and $D$ as follows,
\begin{align}
    V^{\rm SC}=P,\ \ V^{\rm SDW}=-C,\ \ V^{\rm CDW}=2D-C.
\end{align}
In a given channel, the matrix can be decomposed by singular value decomposition (SVD), in momentum space,
\begin{align}
V_{mn}(\0q)=\sum_{\alpha} \phi_{\alpha m}(\0q) S_{\alpha}(\0q) \phi^*_{\alpha n}(\0q),
\end{align}
where $m,n$ label the fermion bilinear, $S_\alpha$ and $\phi_{\alpha n}$ are eigen value and vector for the $\alpha$-th singular mode.

During the SM-FRG flow, we monitor the leading (most negative) eigenvalue, which we abbreviate as $S$ in each channel.
As the energy scale $\Lambda$ reduces, the first divergence of $S$ indicates a tendency towards an instability with order parameter described by the associated eigen mode $\phi(\0Q)$, where $\0Q$ is the associated collective momentum. In this case, one can drop the nonsingular components to write the renormalized interaction as,
\begin{equation}
H_\Gamma \sim \frac{S}{N}O^\dagger O + \cdots,
\end{equation}
where $N$ is the number of unitcells, $O$ is the mode operator that is a combination of the fermion bilinears (see below), and the dots represent symmetry related terms. For example, if the SC channel diverges first, we have
\begin{equation}
O_{SC}^\dagger  = \sum_n \phi_n(\0Q) \alpha_n^\dagger(\0Q) \to  \sum_{\0k,n=(a,b,\delta)}\psi_{\0k+\0Q, a}^\dagger \phi_n(\0Q) e^{i\0k\cdot\delta}\psi_{-\0k,b}^\dagger,\end{equation}
where $n$ labels a fermion bilinear, $a$ and $b$ denote the orbital (and sublattice), and $\Delta_\delta^{ab}=\phi_n(\0Q)$ is the element of the real-space pairing matrix on the bond $\delta$ radiating from $a$.
The spin indices do not have to be specified, as the symmetry of the gap function under inversion automatically determines whether the pair is in the singlet or triplet state.
Because of the eventual Cooper mechanism, the divergence in the pairing channel always occurs at momentum $\0Q=0$.

Similarly, if the SDW channel diverges first, we obtain the mode operator
\begin{equation}
O_{SDW}^\dagger  = \sum_n \phi_n(\0Q)\beta_n^\dagger(\0Q)\to
\sum_{\0k,n=(a,b,\delta)}\psi_{\0k+\0Q,a,\uparrow}^\dagger \phi_n(\0Q) e^{i\0k\cdot\delta}\psi_{\0k,b,\downarrow},\end{equation}
where we assign the spin order in the transverse direction.
Finally, if the CDW channel diverges first, we obtain the mode operator
\begin{equation}
O_{CDW}^\dagger = \sum_n \phi_n(\0Q)\gamma_n^\dagger(\0Q)\to
\sum_{\0k,\sigma,n=(a,b,\delta)}\psi_{\0k+\0Q,a,\sigma}^\dagger \phi_n(\0Q) e^{i\0k\cdot\delta}\psi_{\0k,b,\sigma}.\end{equation}
Note that $O_{SDW/CDW}$ can capture both onsite and on-bond density waves, since the fermion bilinears contain both cases of $\delta=0$ and $\delta\neq 0$.

\subsection{Benchmark of fermion bilinear truncation}

As one of the main approximations in SM-FRG, we impose a truncation length $L_c$ to the internal relative distance between two fermions within a bilinear. For the same vertex $\Gamma_{1234}$ in Fig.S1(a), we require
\begin{align}
|\0R_2-\0R_1|\leq L_c, \ \ |\0R_3-\0R_4|\leq L_c, \ \ \text{for P},\nn\\
|\0R_3-\0R_1|\leq L_c, \ \ |\0R_2-\0R_4|\leq L_c, \ \ \text{for C}, \nn\\
|\0R_4-\0R_1|\leq L_c, \ \ |\0R_2-\0R_3|\leq L_c. \ \ \text{for D}.
\end{align}
We should emphasize that there is no cutoff imposed on the distance between two fermion bilienars in any given channel, and this is important to access the thermodynamic limit. If the four indices of a vertex satisfy more than one of the above truncations, the vertex is said to have overlaps in the channels. In the main text, we choose the truncation length $L_c=\sqrt{2}$, for which up to the 4th nearest neighbors are included in the fermion bilinear basis. (We set the inter-layer distance as $1/2$.)
In Fig.~\ref{fig:Lc}, we present our SM-FRG results for $L_c=0.5$ and $\sqrt{2}$ with the same parameters of Fig.~2(a) in the main text. Clearly, they give almost the same flows. Only slight deviations are seen at low energy scales. The convergence of our results with increasing $L_c$ justifies the truncation approximation.

\begin{figure}[h]
	\includegraphics[width=0.6\textwidth]{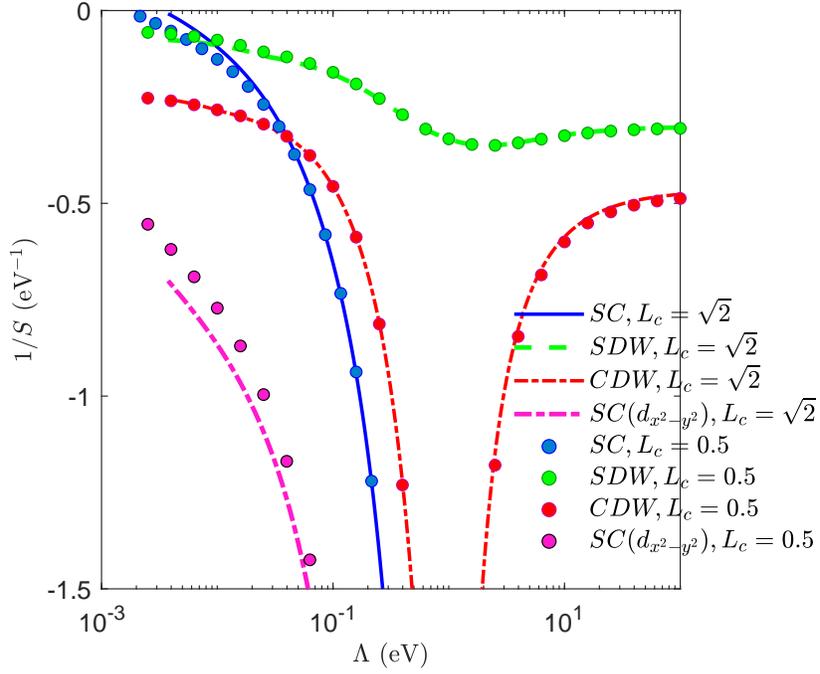}
	\caption{SM-FRG results for the truncation length $L_c=0.5$ and $L_c=\sqrt{2}$, with the same parameters of Fig.~2(a) in the main text.}
	\label{fig:Lc}
\end{figure}

\section{Effective Hamiltonian under the Gutzwiller approximation}
In the main text, we have obtained the effective Hamiltonian $\tilde{H}$ [Eq.(3) in the main text] in the low energy subspace with one or two electrons per site. In this Supplementary Material, we study this effective Hamiltonian under the Gutzwiller approximation \cite{Gutzwiller_PRL_1963,ZhangFC1988} in the {\it paramagnetic and translation-invariant} case since we are mainly interested in superconductivity in this work. Generalizations to other cases are straightforward.

The trial ground state $\ket{G}$ is constructed as $\ket{G}=\hat{P}\ket{0}$ (up to normalization) where $\ket{0}$ is the ground state of a variational Hamiltonian. The projection operator is
\begin{align}
\hat{P}=\prod_i \hat{P}(i),\quad \hat{P}(i)=\eta_1(i)\hat{Q}_1(i)+\eta_2(i)\hat{Q}_2(i) ,
\end{align}
where $\hat{Q}_n(i)$ ($n=1,2$) is the projection operator for $n$-electrons at $i$-site, and $\eta_2=\eta_1^2$ are fugacities introduced to control the total charge, or average filling. Note we generalized the projection operator in the canonical ensemble (defined in the main text) for that in the grand-canonical ensemble, where the derivation is much easier \cite{Wang_PRB_2006,Wang_PRB_2022}. In principle, the fugacity can depend on position, but for our purpose it can be set to be uniform. In the following, for brevity we will suppress the site label $i$ unless confusion would arise.
The projection operator $\hat{Q}_n$ can be further expressed as
\begin{align}
\hat{Q}_1&=\hat{Q}_{1,x}\hat{Q}_{0,z}+\hat{Q}_{0,x}\hat{Q}_{1,z} , \\
\hat{Q}_2&=\hat{Q}_{2,x}\hat{Q}_{0,z}+\hat{Q}_{1,x}\hat{Q}_{1,z}+\hat{Q}_{0,x}\hat{Q}_{2,z} ,
\end{align}
where $\hat{Q}_{n,x/z}$ is the orbital resolved projection operator for $n$-electrons on $x/z$-orbital ($x$ for $d_{x^2-y^2}$ and $z$ for $d_{3z^2-r^2}$ at $i$-site. Spin is not resolved for the paramagnetic case in this study. To be specific,  $\hat{Q}_{n,a}$ ($n=0,1,2; a=x,z$) are defined as
\begin{align}
\hat{Q}_{0,a}&=\left(1-\hat{n}_{a\up}\right)\left(1-\hat{n}_{a\dn}\right) ,\\
\hat{Q}_{1,a}&=\hat{n}_{a\up}\left(1-\hat{n}_{a\dn}\right) + \hat{n}_{a\dn}\left(1-\hat{n}_{a\up}\right) ,\\
\hat{Q}_{2,a}&=\hat{n}_{a\up}\hat{n}_{a\dn} ,
\end{align}
which are empty, single and double occupation operators for the $a$-orbital.

To proceed, we define the expectation values of these projection operators in the unprojected state $\ket{0}$ as
\begin{align}
 q_{n}^{(0)}=\av{\hat{Q}_{n}}_0,\quad q_{n,x/z}^{(0)}=\av{\hat{Q}_{n,x/z}}_0,\quad n=1,2 .
\end{align}
In practice, we replace the operators $\hat{n}_{a}$ by their expectation values $n_{a}^{(0)}$ in the unprojected state $\ket{0}$.
% under the Hartree approximation.

On the other hand, in the projected state $\ket{G}=\hat{P}\ket{0}$, we have
\begin{align}
q_{n}=\av{\hat{Q}_n}_G=\frac{\av{\hat{P}\hat{Q}_{n}\hat{P}}_0}{\av{\hat{P}\hat{P}}_0}=\frac{\eta_{n}^2q_n^{(0)}}{Z}, \quad n=1,2; \quad Z=\eta_{1}^2q_1^{(0)}+\eta_{2}^2q_2^{(0)},
\end{align}
such that
$q_1+q_2=1$ and $\eta_n^2/Z=q_n/q_n^{(0)}$.
With these relations, the expectation values of the orbital resolved projectors on $\ket{G}$ are obtained as
\begin{align}
q_{0,a} &=  \frac{\av{\hat{P}\hat{Q}_{0,a}\hat{P}}_0}{\av{\hat{P}\hat{P}}_0} = \frac{ \av{\eta_1^2\hat{Q}_{0,a}\hat{Q}_{1,\bar{a}}+\eta_2^2\hat{Q}_{0,a}\hat{Q}_{2,\bar{a}}}_0 }{Z}  =q_{0,a}^{(0)} \left(\frac{q_1}{q_1^{(0)}}q_{1,\bar{a}}^{(0)} + \frac{q_2}{q_2^{(0)}}q_{2,\bar{a}}^{(0)}\right) ,\\
q_{1,a}&=\frac{\av{\hat{P}\hat{Q}_{1,a}\hat{P}}_0}{\av{\hat{P}\hat{P}}_0} = \frac{ \av{\eta_1^2\hat{Q}_{1,a}\hat{Q}_{0,\bar{a}}+\eta_2^2\hat{Q}_{1,a}\hat{Q}_{1,\bar{a}}}_0 }{Z} = q_{1,a}^{(0)}\left( \frac{q_1}{q_1^{(0)}} q_{0,\bar{a}}^{(0)} + \frac{q_2}{q_2^{(0)}} q_{1,\bar{a}}^{(0)} \right) ,\\
q_{2,a}&=\frac{\av{\hat{P}\hat{Q}_{2,a}\hat{P}}_0}{\av{\hat{P}\hat{P}}_0} =\frac{ \av{\eta_2^2\hat{Q}_{2,a}\hat{Q}_{0,\bar{a}}}_0 }{Z} = q_{2,a}^{(0)}\frac{q_2}{q_2^{(0)}} q_{0,\bar{a}}^{(0)} ,
\end{align}
where $\bar{a}$ represents the orbital other than $a$.
In the above calculations, we have used Gutzwiller approximation by neglecting inter-site correlations.
From the above results, $q_n$ and $q_{n,x/z}$ are determined by $\eta_n$, $q_{n}^{(0)}$ and $q_{n,x/z}^{(0)}$, while the latter two are given by $\ket{0}$. We tune the fugacity such that
\begin{align}
\av{\hat{Q}_1+2\hat{Q}_2}_G=q_1+2q_2=\frac32 .
\end{align}
Together with the normalization condition $q_1+q_2=1$, we obtain $q_1=q_2=1/2$. Therefore, the variational parameters coming from the projector are determined once  $n_{a\sigma}^{(0)}=n_a^{(0)}/2$ is fixed by the unprojected state $\ket{0}$.

To determine $|0\rangle$, we need to evaluate the total energy $E=\av{\tilde{H}}_G$.
The onsite energy term is
\begin{align}
E_\varepsilon=N\sum_{a=x,z}\epsilon_a (q_{1,a}+2q_{2,a}) ,
\end{align}
where $N$ is the number of sites.
The Hubbard $(U+J_H)$-term ($J_H$ comes from the doublon energy contributed by the Hund's coupling term) is
\begin{align}
E_U=N(U+J_H)(q_{2,x}+q_{2,z}) .
\end{align}
The $U'$-term ($U'=U-2J_H$) is
\begin{align}
E_{U'}=N(U-2J_H)q_{1,x}q_{1,z} .
\end{align}
The above two terms can be added together as $E_U+E_{U'}=NJ_H(q_{2,x}+q_{2,z}-2q_{1,x}q_{1,z})+Nq_2U$. Note $q_2U=U/2$ is a constant and can be dropped.
The Hund's coupling $J_H$- and pair hopping $J_P$-terms ($J_P=J_H$) contribute the energy
\begin{align}
E_{J_H}=\sum_i\underbrace{\frac{q_2}{q_2^{(0)}}}_{g_{H}}J_H\av{\frac14n_{ix}n_{iz}-\0S_{ix}\cdot\0S_{iz} + B_{ix}^\dag B_{iz} +B_{iz}^\dag B_{ix} }_0 ,
\end{align}
where $n_{ia}$, $\0S_{ia}$ and $B_{ia}=c_{ia\dn}c_{ia\up}$ are charge, spin and intra-orbital pairing operators for $a$-orbital at $i$-site. We use the standard mean field decoupling to write
\begin{align}
E_{J_H}=\sum_i g_{H}J_H\left( \frac38\chi_{ixz}^*\chi_{ixz}+\frac38 \Delta_{ixz}^* \Delta_{ixz}+\Delta_{ix}^*\Delta_{iz}+\Delta_{ix}\Delta_{iz}^* \right) ,
\end{align}
where $\chi_{ixz}=\sum_\sigma\av{c_{ix\sigma}^\dag c_{iz\sigma}}_0$, $\Delta_{ixz}=\av{c_{ix\dn}c_{iz\up}-c_{ix\up}c_{iz\dn}}_0$, $\Delta_{ix}=\av{c_{ix\dn}c_{ix\up}}_0$.
Similarly, the intra-orbital super-exchange term between $i$- and $j$-sites becomes
\begin{align}
E_J=\sum_{\av{ij}a} g_J^aJ_{ij}^{a}\av{ \0S_{ia}\cdot\0S_{ja}-\frac14n_{ia}n_{ja} }_0 = -\sum_{\av{ij}a}\frac38 g_J^aJ_{ij}^{a}\left(\chi_{ij}^{a,*}\chi_{ij}^{a}+\Delta_{ij}^{a,*}\Delta_{ij}^{a}\right) ,
\end{align}
where
\begin{align}
g_J^a=\left(\frac{q_1}{q_1^{(0)}}q_{0,\bar{a}}^{(0)}\right)^2+\left(\frac{q_2}{q_2^{(0)}}q_{1,\bar{a}}^{(0)}\right)^2 ,
\end{align}
and $\chi_{ij}^a=\sum_\sigma\av{c_{ia\sigma}^\dag c_{ja\sigma}}_0$, $\Delta_{ij}^a=\av{c_{ia\dn}c_{ja\up}-c_{ia\up}c_{ja\dn}}_0$.
For the hopping term, we obtain
\begin{align}
E_k=\sum_{ijab\sigma}\frac{\av{\hat{P}c_{ia\sigma}^\dag c_{jb\sigma}\hat{P}}_0}{\av{\hat{P}\hat{P}}_0}&=\sum_{ijab\sigma} \frac{\eta_2(i)\eta_1(i)\eta_1(j)\eta_2(j)}{Z(i)Z(j)} \av{\hat{Q}_{1,\cancel{a\sigma}}(i)c_{ia\sigma}^\dag c_{jb\sigma} \hat{Q}_{1,\cancel{b\sigma}}(j)}_0 \nn\\
&=\sum_{ijab\sigma}\underbrace{\frac{q_1q_2}{q_1^{(0)}q_2^{(0)}} q_{1,\cancel{a\sigma}}^{(0)}q_{1,\cancel{b\sigma}}^{(0)}}_{g_t^{ab}}\av{c_{ia\sigma}^\dag c_{jb\sigma}}_0 ,
\end{align}
where $\hat{Q}_{n,\cancel{a\sigma}}(i)$ is the projector for $n$-electrons at $i$-site by ignoring $a$-orbital and $\sigma$-spin (which are automatically taken care of by the electron annihilation and creation operators). Its expectation value in $\ket{0}$ is
\begin{align}
q_{1,\cancel{a\sigma}}^{(0)}=n_{a\bar{\sigma}}^{(0)}q_{0,\bar{a}}^{(0)}+\left(1-n_{a\bar{\sigma}}^{(0)}\right)q_{1,\bar{a}}^{(0)} .
\end{align}

Combining all ingredients, we get the total energy $E=E_k+E_U+E_{U'}+E_{J_H}+E_{J}+E_{\varepsilon}$ as \begin{align}
E=&\sum_{\av{ij}ab}g_t^{ab}\left(t_{ij}^{ab}\chi_{ij}^{ab}+h.c.\right) - \sum_{\av{ij}a}\frac38 g_J^a J_{ij}^a \left( \chi_{ij}^{a,*}\chi_{ij}^a + \Delta_{ij}^{a,*}\Delta_{ij}^a\right) \nn\\
&+\sum_{i}g_{H}J_H\left(\frac38\chi_{xz}^*\chi_{xz}+\frac38 \Delta_{xz}^* \Delta_{xz}+\Delta_x^*\Delta_z+\Delta_x\Delta_z^*\right) \nn\\
&+NJ_H\left(q_{2,x}+q_{2,z}-2q_{1,x}q_{1,z}\right) + N\sum_{a} \epsilon_a\left(q_{1,a}+2q_{2,a}\right).
\end{align}
Then, by optimizing $E$ with respect to $\bra{0}$,
under the normalization condition $\av{0|0}=1$ and $\av{n_{ia}}_0=n_{a}^{(0)}$, we obtain the condition $(H_v-\lambda)\ket{0}= 0$, with the variational Hamiltonian $H_v$ given by
\begin{align}
H_{v}=&\sum_{\av{ij}ab\sigma}g_t^{ab}\left( t_{ij}^{ab} c_{ia\sigma}^\dag c_{jb\sigma} + h.c.\right) -\sum_{ia\sigma}\mu_a c_{ia\sigma}^\dag c_{ia\sigma} \nn\\
&- \sum_{\av{ij}a\sigma} \frac38 g_J^a J_{ij}^a \left(\chi_{ij}^{a,*}c_{ia\sigma}^\dag c_{ja\sigma} + \Delta_{ij}^{a,*}\sigma c_{ia\bar{\sigma}}c_{ja\sigma}+h.c.\right) \nn\\
&+\sum_{i\sigma} g_H J_H\left(\frac38\chi_{xz}^*c_{ix\sigma}^\dag c_{iz\sigma} + \frac38\Delta_{xz}^* \sigma c_{ix\bar{\sigma}}c_{iz\sigma} + \frac12\Delta_x^*\sigma c_{iz\bar{\sigma}}c_{iz\sigma} + \frac12\Delta_z^*\sigma c_{ix\bar{\sigma}}c_{ix\sigma} + h.c.\right) ,
\end{align}
where $\sigma=\pm1$ for spin up/down, and $\mu_a$ is the Lagrangian multiplier to enforce $\av{n_{ia}}_0=n_{a}^{(0)}$. This is the detailed form of $H_v$ resorted to in the main text.

In practice, we solve this variational problem in two steps: (i) For given ($n_{x}^{(0)}$, $n_{z}^{(0)}$) (with only one free parameter due to $n_x^{(0)}+n_z^{(0)}=3/2$), all the renormalized factors ($g_t^{ab}$, $g_J^a$, $g_H$) are obtained directly. We solve $\Delta$'s and $\chi$'s self-consistently, and simultaneously tune $\mu_{x/z}$ to match the unprojected density $n_{x/z}^{(0)}=\av{n_{x/z}}_0$. After the self-consistency is achieved, we calculate $E$. (ii) We optimize $E$ with respect to $n_x^{(0)}$, \ie $\partial E/\partial n_x^{(0)}=0$.
The optimization process for $(U,J_H)=(8,0.3)$~eV is shown in Fig.~\ref{fig:rmft-sm} for (a) the energy $E$, and (b) the corresponding pairing components $\{u_{0,1}, v_{0,2}\}$ (see definitions in the main text).

\begin{figure}[h]
\includegraphics[width=0.6\linewidth]{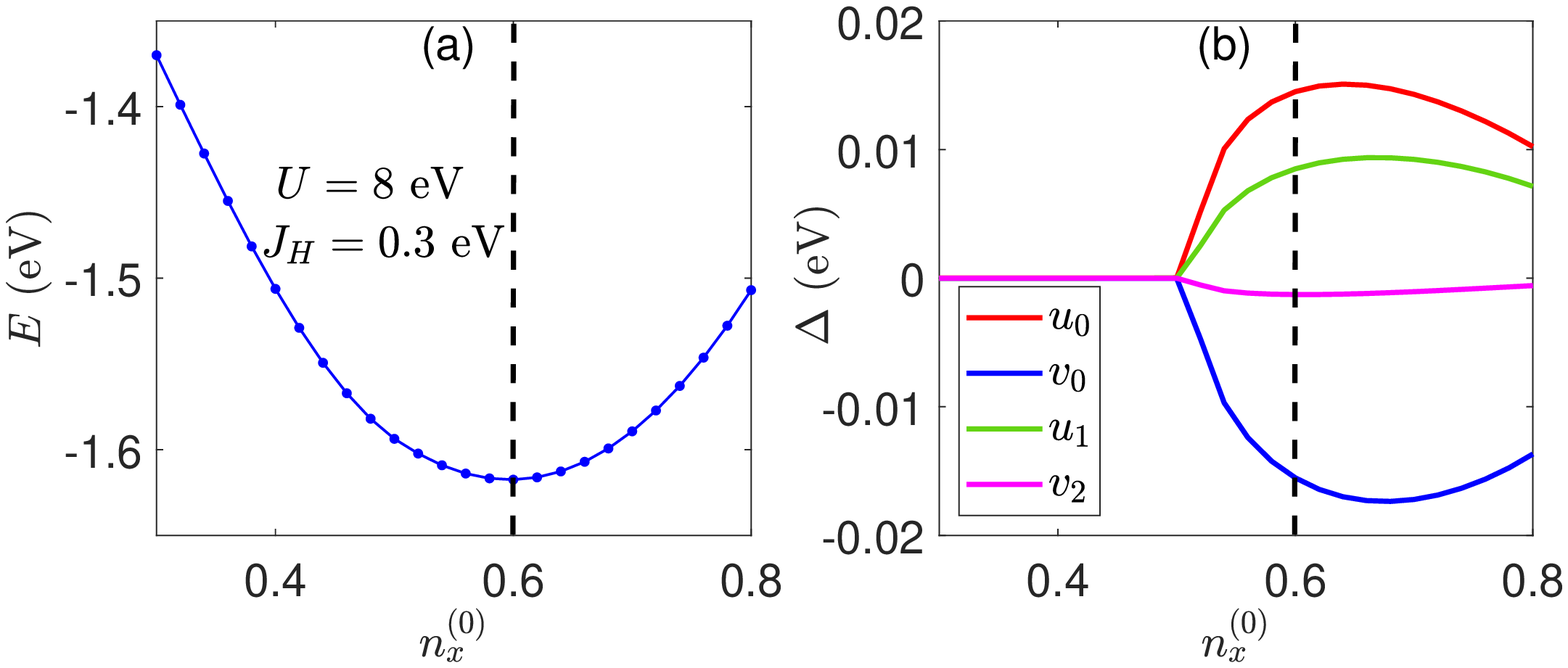}
\caption{\label{fig:rmft-sm}
(a) The energy $E$ versus the variational parameter $n_{x}^{(0)}$ for $(U,J_H)=(8,0.3)$~eV. (b) The corresponding pairing amplitudes $\{u_{0,1}, v_{0,2}\}$ (see definitions in the main text) versus $n_{x}^{(0)}$. The dashed lines highlight the optimized value of $n_{x}^{(0)}$.
}
\end{figure}
%\bibliography{La3Ni2O7}
\end{widetext}

\end{document}